\documentclass[]{sig-alternate-10pt}
\pdfoutput=1

\usepackage{tikz}
\usepackage{amsmath}

\usepackage{blindtext}
\usepackage[normalsize,TABBOTCAP]{subfigure}
\usepackage{footmisc}

\usepackage{filecontents}
\usepackage{algpseudocode}
\usepackage{xspace}

\usepackage{verbatim}
\usepackage{color}
\usepackage{verbatim}
\usepackage{url}
\usepackage{graphicx}
\graphicspath{
	{./figure}
}
\usepackage{balance}
\usepackage{hyperref}
\usepackage{color}
\usepackage{comment}
\usepackage{makecell}
\usepackage{caption}

\usepackage{amsmath}
\usepackage{algorithm}
\usepackage{array}
\usepackage[toc,page]{appendix}
\usepackage{xspace}
\usepackage{xfrac}
\usepackage{amssymb}
\usepackage{mathrsfs}
\usepackage{wrapfig}
\usepackage[blocks]{authblk}

\usepackage{amsmath}

\begin{document}
    \title{mmWave Wearable Antenna for Interaction with VR Devices}

\author[ ]{\Large{\textsf{Haksun Son, Song Min Kim
}}}
\setlength{\affilsep}{0.1em}
\affil[ ]{\normalsize{Korea Advanced Institute of Science and Technology (KAIST)}}
\affil[ ]{\normalsize{\{sonhaksun10, songmin\}@kaist.ac.kr}}

\maketitle
  	\begin{abstract}
The VR industry is one of the most promising industries for the near future, as it can provide a more immersive connection between people and the virtual world. Currently, VR devices interact with people using inconvenient controllers or cameras that perform poorly in dark environments. Interaction through millimeter-wave wearable devices has the potential to conveniently track human behavior regardless of the lighting conditions. In this study, a millimeter-wave wearable antenna was developed, opening up the possibility for more immersive interaction with VR devices. The antenna features a low loss tangent polyester fabric to minimize dielectric losses and a smooth coating to reduce losses due to rough surfaces. The antenna operates in the 24GHz ISM band, with an S11 value of -29dB at 24.15GHz.
\end{abstract}
  	
\section{Introduction}
Recent advancements in augmented and extended reality (AR/VR) devices hold promise for providing various realistic services including sports science, physical therapy, and gaming applications. The extended reality market is projected to grow by 24.2 percent annually, reaching a market size of 1,246 billion US dollars by 2035~\cite{vrtrend}.

Current VR devices use controllers to interact with people or use cameras to recognize actions. 
However, controllers are inconvenient for everyday use, and cameras do not work well in dark environments, disrupting immersive interaction. In this work, we utilize RF-based approaches, which have long been studied with WiFi or radars~\cite {jeong2020sdr, kong2022m3track} to achieve non-intrusive, and seamless VR.

Wearable devices are designed to be comfortable for users and, using RF signals, can operate well regardless of ambient lighting.
Wearable devices for VR require communication with multiple users and multiple devices simultaneously and also require precise location detection for interaction based on user action and location-based services. In millimeter-wave technology, a wide bandwidth can be utilized, making it possible to communicate simultaneously with many users and achieve precise location detection. With a 250MHz bandwidth in the 24GHz band, simultaneous communication with 1100 people is possible, and device positioning with an error of within 1cm is achievable~\cite{bae2022omniscatter, bae2023hawkeye}. 

In this study, a wearable millimeter-wave antenna for VR devices was developed. Wearable antennas operating in the millimeter-wave range are made based on fabric. However, these antennas suffer from strong signal loss due to dielectric losses and losses caused by rough surfaces~\cite{liu2023surface, wang2015embroidered}. In this research, to reduce losses due to dielectrics, polyester with a low loss tangent was used, and to minimize losses due to rough surfaces, a smooth coating was added to the rough fabric. The antenna can operate in the 24GHz ISM band (24GHz – 24.25GHz), with an S11 value of -29dB at 24.15GHz.

  	\begin{table}[h!]
\centering
\begin{tabular}{|c|c|c|}
\hline
Fabric    & Permittivity & Loss tangent \\ \hline
Silk      & 1.75         & 0.012        \\ \hline
Felt      & 1.38         & 0.023        \\ \hline
Cotton    & 1.6          & 0.0098       \\ \hline
Polyester & 1.9          & 0.0045       \\ \hline
\end{tabular}
\captionof{table}{Permittivity and loss tangent of various fabrics~\cite{baiya2014development}.} 
\label{table:permittitivy}
\end{table}

\begin{figure}[h!]
\centering
\vspace{-2mm}
\includegraphics[width=\columnwidth]{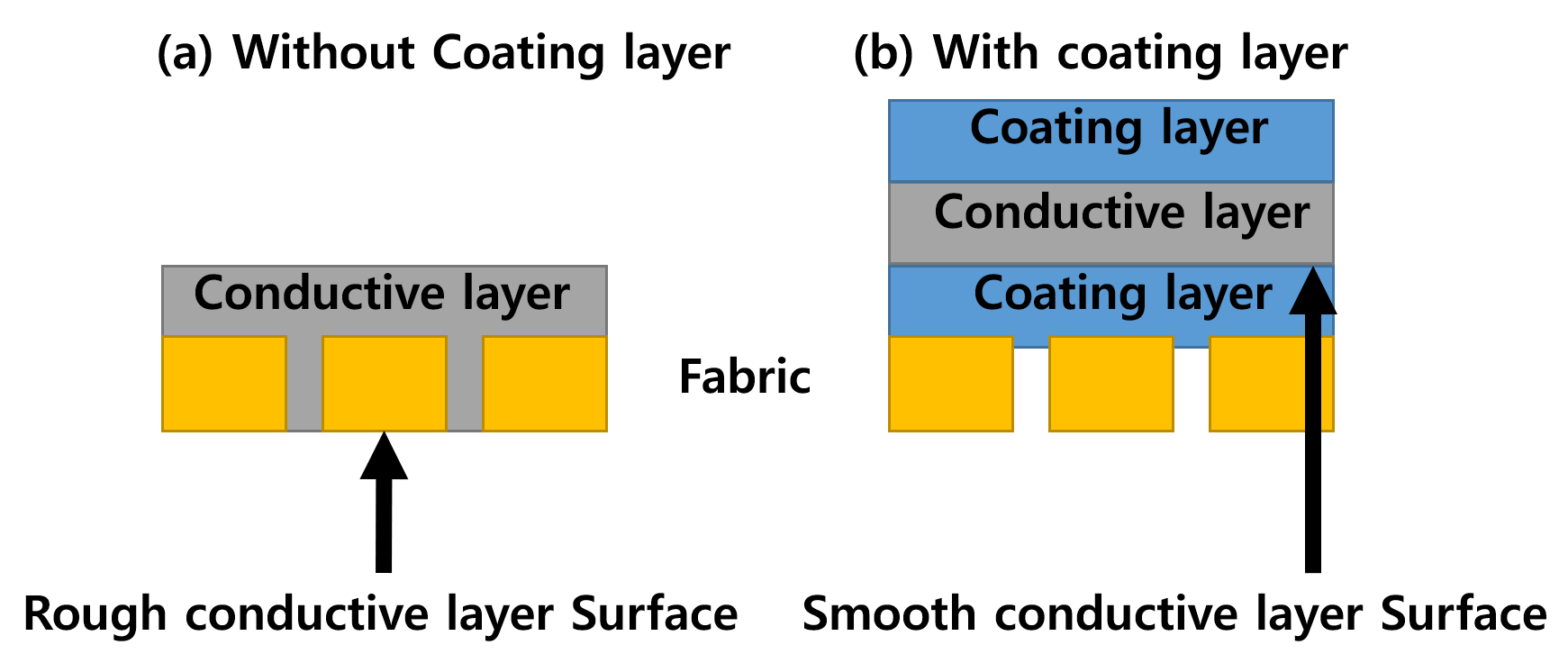}
\vspace{-2mm}
\caption{Layer structure of the antenna with and without coating. (a) Without coating (b) With coating.
}
\label{fig:layer}
\vspace{-0.4cm}
\end{figure}

\section{Design}\label{sec:tag}
\subsection{Wearable Antenna Design}
Tags attached to various parts of a person's body to track their movements should be made from fabric that is convenient to wear. In this study, we used polyester fabric among various commonly used fabrics. Referring to Table 1, polyester fabric has a lower loss tangent compared to other fabrics, and it can reduce dielectric loss inside the transmission line~\cite{baiya2014development}.

Various technologies have been researched for creating antennas on fabric, including conductive thread, conductive cloth, and screen printing~\cite{wang2015embroidered}. In this study, we used conductive ink and a coating paper, which can minimize the signal loss caused by a rough surface. Additionally, production with a printer can reduce the production cost and enable many people to use it easily.

Figure 1 depicts the structure of each layer of the antenna. There are three major issues when there is no coating. First, due to the rough surface of the fabric, a conductive layer forms, causing strong signal loss~\cite{liu2023surface}. Second, if the conductive ink flows down to the ground, the antenna pattern can be grounded, leading to a complete loss of antenna functionality. The last issue is that the conductive layer is easily damaged by wear and tear. To address these issues, we added a coating layer on both sides of the conductive layer. The first coating layer is intended to smooth out the conductive layer to reduce signal loss and prevent the conductive ink from leaking to the ground. Another coating layer protects from wear and tear caused by external forces.

We used polyester fabric with a thickness of 0.17mm. Additionally, we used the NBSIJ-MU01, a conductive ink compatible with a household printer, and used transparent silhouette temporary tattoo paper as the coating paper. The production process is as follows:

\vspace{-1mm}
\begin{enumerate}
  \item Apply a layer of coating paper onto the polyester.
  \item Use a printer filled with conductive ink to print the RF circuit onto another layer of coating paper.
  \item Sprinkle saltwater containing chlorine ions to clump the silver nanoparticles in the conductive ink.
  \item Apply another layer of coating paper over the polyester fabric with the RF circuit drawn on it.
  \item Repeat steps 1-4 on the backside of the polyester fabric or add a ground using copper tape.
\end{enumerate}

\begin{figure}[h!]
\centering
\vspace{-2mm}
\includegraphics[width=\columnwidth]{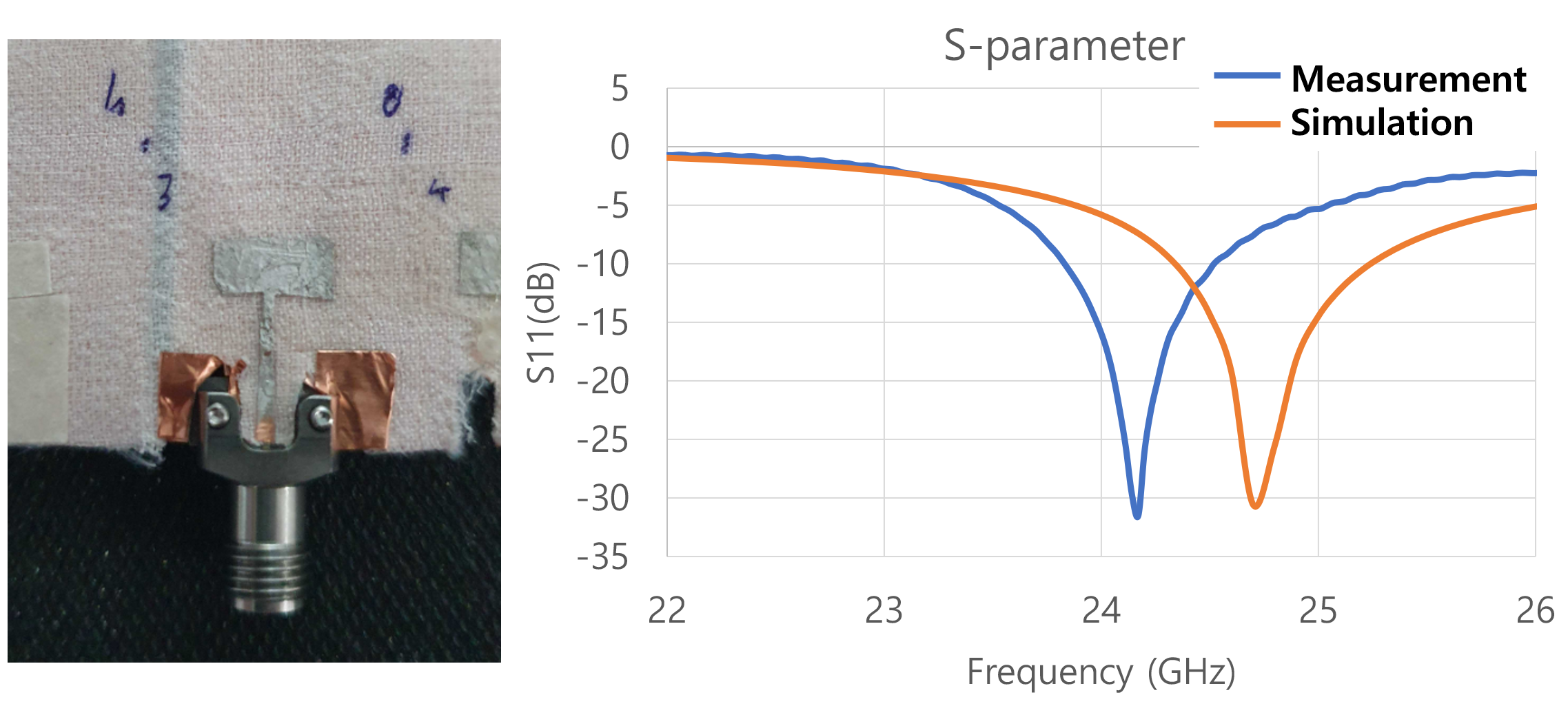}
\vspace{-2mm}
\caption{Appearance of wearable antenna and antenna’s S-parameter (Measurement and Simulation)
}
\label{fig:performance}
\vspace{-0.4cm}
\end{figure}

\subsection{Performance of Wearable Antenna}
Figure 2 shows an antenna using polyester fabric and conductive ink. The dimensions of this antenna are 3.9mm in height and 8.8mm in width. The S-parameters of this antenna showed a shift of approximately 520MHz between the results from the simulation and the actual measured values. The S11 was maximum at –29dB at 24.15GHz, and the S11 values were below -10dB in the frequency range from 23.8GHz to 24.5GHz. This indicates that the antenna operates well in the 24GHz ISM band (24GHz – 24.25GHz). 
\balance
\bibliographystyle{abbrv}

\bibliography{ref}  

\begin{thebibliography}{1}

\bibitem{bae2022omniscatter}
K.~M. Bae, N.~Ahn, Y.~Chae, P.~Pathak, S.-M. Sohn, and S.~M. Kim.
\newblock Omniscatter: extreme sensitivity mmwave backscattering using commodity fmcw radar.
\newblock In {\em MobiSys' 22}, pages 316--329, 2022.

\bibitem{bae2023hawkeye}
K.~M. Bae, H.~Moon, S.-M. Sohn, and S.~M. Kim.
\newblock Hawkeye: Hectometer-range subcentimeter localization for large-scale mmwave backscatter.
\newblock In {\em MobiSys' 23}, pages 303--316, 2023.

\bibitem{baiya2014development}
D.~Baiya.
\newblock On the development of conductive textile antennas.
\newblock 2014.

\bibitem{vrtrend}
{EMERGEN Research}.
\newblock {Extended Reality Market Size, Share, Trends, By Type (Business Engagement, Customer Engagement), By Application (Virtual Reality (VR), Augmented Reality (AR), Mixed Reality (MR)), By End-use (BFSI, Education, Consumer Goods \& Retail, Industrial \& Manufacturing), and By Region Forecast to 2035.}
\newblock https://www.emergenresearch.com/industry-report/extended-reality-market.

\bibitem{jeong2020sdr}
W.~Jeong, J.~Jung, Y.~Wang, S.~Wang, S.~Yang, Q.~Yan, Y.~Yi, and S.~M. Kim.
\newblock Sdr receiver using commodity wifi via physical-layer signal reconstruction.
\newblock In {\em MobiCom' 20}, pages 1--14, 2020.

\bibitem{kong2022m3track}
H.~Kong, X.~Xu, J.~Yu, Q.~Chen, C.~Ma, Y.~Chen, Y.-C. Chen, and L.~Kong.
\newblock m3track: mmwave-based multi-user 3d posture tracking.
\newblock In {\em MobiSys' 22}, pages 491--503, 2022.

\bibitem{liu2023surface}
Y.~Liu, Y.~Guo, C.~Li, X.~Ye, and D.~Kim.
\newblock Surface roughness effect from different surfaces of microstrip lines and reference plane.
\newblock {\em IEEE Letters on Electromagnetic Compatibility Practice and Applications}, 2023.

\bibitem{wang2015embroidered}
Z.~Wang, J.~Volakis, and A.~Kiourti.
\newblock Embroidered antennas for communication systems.
\newblock In {\em Electronic textiles}, pages 201--237. Elsevier, 2015.

\end{thebibliography}

\end{document}